\documentclass{article}
\usepackage[utf8]{inputenc}
\usepackage{tabularx} 
\usepackage{amsmath}  
\usepackage{listings}
\usepackage[margin=1in,letterpaper]{geometry} 
\usepackage{graphicx}
\usepackage{dcolumn}
\usepackage{bm}
\usepackage{hyperref}
\hypersetup{
colorlinks = true,
linkcolor = blue,
citecolor = blue,
filecolor = magenta,
urlcolor = blue
} 
\usepackage[mathlines]{lineno}
\usepackage{amsmath}
\usepackage{amssymb}
\DeclareMathOperator{\tr}{tr}
\usepackage{physics}
\usepackage{soul}
\usepackage[labelfont=bf,labelsep=quad,justification=justified]{caption}
\usepackage{subcaption}
\usepackage{dcolumn}
\usepackage{bm}
\usepackage[english]{babel}
\usepackage[T1]{fontenc}
\usepackage[utf8]{inputenc}
\usepackage{xcolor}
\usepackage{float}
\usepackage[style=numeric-comp,sorting=none]{biblatex}
\usepackage{romannum}

\usepackage{authblk}

\begin{document}

\title{Influence of the coherence of spectral domain interference of Fano resonance on the degree of polarization of light}

\author[1]{Shyamal Guchhait}
\author[3]{Devarshi Chakrabarty}
\author[3]{Avijit Dhara}
\author[1]{Ankit Kumar Singh}
\author[3]{Sajal Dhara}
\author[1,4]{ Nirmalya Ghosh}
\affil[1]{Department of Physical Sciences (DPS) \\ 

Indian Institute of Science Education and Research (IISER) Kolkata \\ 

Mohanpur, India - 741 246}
\affil[3]{Department of Physics \\ 

Indian Institute of Technology Kharagpur \\ 

India - 731202}
\affil[4]{
Center of Excellence in Space Sciences India | CESSI \\

Indian Institute of Science Education and Research (IISER) Kolkata \\ 

Mohanpur, India - 741 246}
\affil[*]{\href{mailto:sg16ip022@iiserkol.ac.in}{sg16ip022@iiserkol.ac.in}, \href{mailto:sajaldhara@phy.iitkgp.ac.in }{sajaldhara@phy.iitkgp.ac.in}}
\date{}
\maketitle

\pagenumbering{arabic}

\begin{abstract}
    We show an intriguing connection between the coherence of spectral domain interference of two electromagnetic modes in Fano resonance and the resulting degree of polarization of light. A theoretical treatment is developed by combining a general electromagnetic model of partially coherent interference of a spectrally narrow and a broad continuum mode leading to Fano resonance and the cross-spectral density matrix of the interfering polarized fields of light. The model suggests a characteristic variation of the degree of polarization across the region of spectral dip and the peak of Fano resonance as an exclusive signature of the connection between the degree of polarization and the coherence of the interfering modes. The predictions of the model is experimentally verified in the partially polarized Fano resonance spectra from metal Chalcogenides systems, which emerged due to the interference of a narrow excitonic mode with the background continuum of scattered light in the reflectance spectra from the system. The demonstrated connection between polarization and coherence in the spectral domain Fano-type interference of electromagnetic modes is fundamentally important in the context of a broad variety of non-trivial wave phenomena that originate from fine interference effects, which may also have useful practical implications.      
\end{abstract}

\section{Introduction\label{introduction}}

Coherence and polarization \cite{cohe.wolf2003unified, cohe.wolf2007introduction,cohe.shirai2004coherence,cohe.shirai2007correlations,cohe.martienssen1964coherence} are two fundamental entities of light which are described through the electric field \cite{em.gori2008realizability,em.korotkova2005generalized} components and their correlations \cite{cohe.martienssen1964coherence}. While the formal definition of degree of polarization (DOP) follows from the so-called coherency matrix, there was no direct connection in terms of their theoretical formalism between the degree of coherence and DOP of any partially coherent light field until recently. Usually DOP and degree of coherence of light fields were treated independently until a connection between these two entities was established through the use of the so-called cross-spectral density matrix formalism \cite{cohe.wolf2003unified, cohe.wolf2007introduction}. The predictions of this model on the connection between DOP and degree of coherence were subsequently experimentally verified by conducting a double-slit interference experiment \cite{cohe.double.li2006spectra,cohe.double.gori2006effects,cohe.double.roychowdhury2005young} using partially polarized, partially coherent light. It was shown that for a partially polarized light, the DOP of the light beam observed in the interference plane depends on the coherence of the two interfering light beams or on the resulting visibility of the interference fringes. As a signature of this connection between DOP and temporal coherence, the DOP of light in the interference plane was observed to vary spatially across the interference minima and the maxima position for a fixed degree of coherence of the interfering beams. Note that interference of waves is a ubiquitous phenomenon observed in various classical and quantum systems dealing with interference of both classical electromagnetic (EM) waves, interference of EM modes,  quantum matter wave, quantum states, etc.  In this regard, there are a rich variety of seemingly counter-intuitive optical wave phenomena that originates due to fine interference effects, e.g., superluminal propagation of wave packets, Fano resonance, Electromagnetically induced transparency (EIT) and absorption (EIA), coherent perfect absorption and super scattering, optical weak value amplification, etc. It is therefore important to ask the question whether such relation between the coherence and DOP can be observed in broader variety of wave phenomena involving interference of waves and  electromagnetic modes. Here, we show that the phenomenon is indeed more general, and it can be observed in the spectral domain interference of two electromagnetic modes of a hybrid system that leads to Fano resonance having asymmetric spectral line shape. We show an intriguing manifestation of the degree of coherence of  two spectrally interfering modes on the spectral variation of DOP of light reflected from a Fano resonant system \cite{fano.kotur2016spectral,fano.christ2003waveguide,fano.miroshnichenko2010fano,fano.ray2017polarization, fano.singh2018tunable,fano.grillet2006characterization,fano.poddubny2012fano,fano.ott2013lorentz}. For this purpose, we combine the electromagnetic model of  Fano resonance involving interference of a discrete, spectrally narrow mode and a spectrally broad mode (or ideal continuum) with the framework of the cross-spectral density matrix of the interfering vector (polarized) fields. We show an interesting correlation between the visibility or the contrast of the spectral domain Fano interference (related to the Fano coherence) and spectral variation of DOP of light scattered from the Fano resonant system. The connection between the DOP and coherence in spectral domain interference is experimentally demonstrated using the Fano resonance in the reflectance spectrum of a transition metal dichalcogenide (TMD) system \cite{chal.gao2013nanostructured,chal.zhou2018library,chal.mose2.tongay2012thermally,chal.ye2015monolayer,chal.lundt2017observation,chal.wse2.lackner2021tunable,chal.mueller2018exciton,chal.schneider2018two,chal.xiao2017excitons}, where Fano resonance emerged from the interference of a spectrally narrow excitonic mode with the background continuum of scattered light. The mutual influence of the degree of coherence on the DOP of the scattered light was vividly manifested in the spectral variation of the DOP around the Fano spectral dip region and was in good agreement with the predictions of the corresponding theoretical model. These findings are expected to bear important consequences both in terms of broad variety of fundamental optical effects that arise due to fine interference and in terms of potential applications. 

\section{Theory\label{theory}}

\textbf{Effect of coherence of the spectral domain interference of Fano resonance on the degree of polarization of light}

The optical Fano resonance can be generally modeled as the spectral domain interference of a spectrally narrow discrete EM mode with a broad continuum mode. The field of the discrete mode can be represented by a complex Lorentzian function $E_1(\omega,t)=\frac{q-i}{\epsilon+i}$ and the corresponding field of an ideal continuum mode is represented as a frequency-independent constant. Here, $\epsilon=\epsilon(\omega)=\frac{\omega-\omega_0}{\gamma/2}$ is the reduced energy where $\omega$ is the frequency of the electromagnetic wave, $\omega_0$, and $\gamma$ are the resonance frequency and the width of the narrow resonance, respectively. $q$ is the so-called Fano asymmetry parameter that determines the asymmetry of spectral line shape. Thus the scattered field  from a Fano resonant system can be written as
\begin{equation}
    \centering
    E(\omega,t) = E_1(\omega,t) + E_2(\omega,t) = \frac{q-i}{\epsilon+i} + e^{i\delta(t)}
    \label{eqs:Fano electric field}
\end{equation}
Here the phase factor $\delta(t)$ includes the stochastic nature of the wave interference.  A completely random distribution of $\delta$ over time ($t$) gives rise to incoherent interference of the two modes, leading to a Lorentzian-type intensity profile having an additional frequency-independent background. On the other hand, in case of ideal Fano resonance  $(\delta(t)\approx0)$ and the corresponding intensity  of Fano resonance is obtained as

\begin{equation}
    \centering
    I(\omega) = \abs{E(\omega)}^2 = \frac{(q+\epsilon)^2}{\epsilon^2+1}
    \label{eqs: Fano intensity}
\end{equation}

Eq. \eqref{eqs: Fano intensity} describes a perfectly coherent spectral domain Fano interference between the two modes. \textbf{\textit{Clearly, destructive interference occurs at the Fano reduced energy $\epsilon=-q$ with corresponding}} frequency $\omega_F=\omega_0-\frac{q\gamma}{2}$, where the intensity becomes zero. On the other hand, maximum intensity is obtained for $\epsilon=\frac{1}{q}$ corresponding to frequency  $\omega_m=\omega_0+\frac{\gamma}{2q}$. Note that the total phase difference between the interfering modes $\psi(\omega)=\tan^{-1}\frac{q+\epsilon}{1-q\epsilon}$. As shown in previous literature, the total phase comprises two parts, one associated with the narrow resonance $(\theta(\omega)=-\tan^{-1}{\frac{1}{\epsilon})}$ and the other related to the Fano phase shift $(\phi_F=-\tan^{-1}{\frac{1}{q}})$, which determines the asymmetry of the line-shape\cite{fano.ray2017polarization,fano.ott2013lorentz}. As expected, at the frequency corresponding to destructive interference  $(\omega_F)$ total phase difference is $\psi=\pi$ .

In the general case of  partial coherence, we have an additional term $(\delta(t))$, which represents the random phase fluctuation and thus includes the effect of decoherence. In such case, the expression for the time-averaged intensity can be obtained as (supporting information)

\begin{equation}
    \centering
    I(\omega) = <E(\omega,t)^*E(\omega,t)>=\frac{(\epsilon+\eta q)^2}{\epsilon^2+1} + \frac{q^2(1-\eta^2)+2(1-\eta)}{\epsilon^2+1}
    \label{eqs: Fano inensity eta}
\end{equation}

Where $\eta$ emerges due to time averaging of randomly fluctuating phase and represents the spectral degree of coherency of Fano interference. Equation 3 shows that the value of $\eta$ can be determined from the experimentally recorded spectral variation of the Fano intensity profile. As apparent, by measuring the intensity as a function of $\omega$ one can estimate the value of $\eta$ and other parameters of Fano resonance ($q$, $\gamma$, $\omega_0$), by fitting it to Eq. \eqref{eqs: Fano inensity eta}. As expected, for the case of perfect coherence  $(\eta=1)$, Eq. \eqref{eqs: Fano inensity eta} reduces to Eq. \eqref{eqs: Fano intensity}. On the other hand, for the case of complete decoherence $(\eta=0)$, the modes do not interfere, and the spectrum becomes a pure Lorentzian with a background. It is thus evident in the intermediate case of partial coherence $(0<\eta<1)$, the line shape would deviate from ideal Fano spectral line shape, specifically the intensity corresponding to the  Fano dip  $(\omega=\omega_F)$ will be non-zero due to an additional Lorentzian background. 

We now proceed to include the effect of polarization in our model through the so-called cross-spectral density matrix $(W)$, which is defined as a Fourier transformation of the electric mutual coherence matrix, which is conventionally defined for space-time representation of coherence\cite{cohe.wolf2007introduction}. As previously mentioned, this formalism was introduced to include the effect of degree of coherence in interference on the resulting DOP of light as\cite{cohe.wolf2007introduction,cohe.shirai2004coherence}
\begin{equation}
    \centering
    W(r_1,r_2,\omega) = 
    \begin{pmatrix}
    W_{xx}(r_1,r_2,\omega) & W_{xy}(r_1,r_2,\omega) \\
    W_{yx}(r_1,r_2,\omega) & W_{yy}(r_1,r_2,\omega)
    \end{pmatrix}
    =[W_{ij}(r_1,r_2,\omega)]_{(i=x,y;j=x,y)} = [<E_i(r_1,\omega)E_j(r_2,\omega)>]_{(i=x,y;j=y,x)}
    \label{eqs:Wolf equation}
\end{equation}
Where $r_1$ \& $r_2$ are two spatial points in the space-frequency representation of the degree of coherence. Here, mutual coherence is calculated between the point $r_1$ \& $r_2$, and $W_{ij}$ is the $ij$-th component of the $W$ matrix, and $E_i$, $E_j$ are the complex electric field components in two orthogonal directions. We now proceed to determine the elements of the $W$ matrix for spectral-domain interference of Fano resonance. For this purpose, we consider two interfering modes whose electric fields are given as  $E^1(\rho_1)$, and $E^2(\rho_2)$, where the symbol $\rho_1$, $\rho_2$ are used in place of $r_1$, $r_2$ to represent the two modes. Note that this is analogous to Wolf's treatment of a double-slit interference, but instead of interference expressed in position space, here interference is expressed in the spectral domain. For a general anisotropic system, the two modes can be represented as\cite{fano.ray2017polarization}
\begin{subequations}
\begin{equation}
    \centering
    E(\rho_1,\rho_2) = E_x(\rho_1) + E_x(\rho_2) = \frac{q_x-i}{\epsilon_x+i} +e^{i\delta_x(t)}
\end{equation}
\begin{equation}
    \centering
    E(\rho_1,\rho_2) = E_y(\rho_1) + E_y(\rho_2) = \frac{q_y-i}{\epsilon_y+i} +e^{i\delta_y(t)}
\end{equation}
\label{eqs:interfering xy modes}
\end{subequations}
Here, $x$ and $y$ are two orthogonal polarization components. For the sake of simplicity (which is also the case in our experiment, described subsequently), we consider the system to be isotropic $(q_x=q_y=q,  \omega_{0x}=\omega_{0y}=\omega_0)$. The corresponding elements of the $W$ matrix of spectral-domain Fano interference can be written as  
\begin{subequations}
    \begin{equation}
        W_{xx/yy} = <E^{*}_{x/y}E_{x/y}> = \frac{q^2+1}{\epsilon^2+1} + 1 + 2\sqrt{\frac{q^2+1}{\epsilon^2+1}} \eta_{xx/yy}\cos{\psi_{x/y}}
    \end{equation}
    \begin{equation}
        \centering
        W_{xy/yx} = <E^{*}_{x/y}E_{y/x}> = \frac{q^2+1}{\epsilon^2+1} + 1 + 2\sqrt{\frac{q^2+1}{\epsilon^2+1}} \eta_{xy/yx}\cos{\psi_{x/y}}
    \end{equation}
    \label{eqs:Fano Wxx/yy}
\end{subequations}
Here $\eta_{xx/yy}$ represents the correlation between the components having the same polarization (x or y) of the electric field of the two interfering modes $E^1(\rho_1)$ and $E^2(\rho_2)$. On the other hand $\eta_{xy/yx}$ represents the correlation between the cross-polarized components (x or y) of the electric fields of $E^1(\rho_1)$ and $E^2(\rho_2)$\cite{cohe.double.li2006spectra}. It follows from the symmetry of the problem $\eta_{xx}=\eta_{yy}$, $\eta_{xy}=\eta_{yx}$. 
The spectral DOP of the stochastic electric fields in the spectral domain Fano interference can then be obtained from the $W$ matrix as\cite{cohe.wolf2007introduction}

\begin{equation}
\centering
    DOP = \sqrt{1-\frac{4\det{W}}{\tr{W}^2}} = \frac{W_{xy/yx}}{W_{xx/yy}}
    \label{eqs:dop jones matrix}
\end{equation} 

Using  Eq. \eqref{eqs:Fano Wxx/yy} the expression for the DOP of the spectral domain interference can finally be obtained as  
\begin{equation}
    \centering
    DOP = \frac{W_{xy}}{W_{xx}} = \frac{\tan{\alpha}\left( \frac{q^2+1}{\epsilon^2+1}+1+2\eta_{xy}\cos{\psi_x}\sqrt{\frac{q^2+1}{\epsilon^2+1}}\right)}{\frac{q^2+1}{\epsilon^2+1}+1+2\eta_{xx}\cos{\psi_x}\sqrt{\frac{q^2+1}{\epsilon^2+1}}}
    \label{eqs:dop W}
\end{equation}

Note that the off-diagonal elements $W_{xy}$ and $W_{yx}$ of the $(W)$ matrix contribute to the depolarization property (or DOP) of the electromagnetic wave \cite{cohe.wolf2007introduction}. Further, the coherence parameters appearing in the $W$ matrix $(\eta_{xx}, \eta_{xy})$ cannot assume arbitrary values. The non-negative semi-definiteness of the $W$ matrix imposes the restriction that $\eta_{xy}>\eta_{xx/yy}$.\cite{cohe.wolf2007introduction, cohe.double.roychowdhury2005young}

In our experiment, we measured the DOP by measuring the Stokes vector elements (which also follows from the elements of the coherency matrix $W$ of Eq. \eqref{eqs:dop W}) as\cite{pola.gupta2015wave}
\begin{equation}
    \centering
    DOP = \frac{\sqrt{Q^2+U^2+V^2}}{I}
    \label{eqs:dop Stokes}
\end{equation}

Here, $I, Q, U, V$ are the first, second, third, and fourth elements of the well-known Stokes vector. 
In what follows, we experimentally demonstrate the role of coherence of spectral domain interference of Fano resonance on the spectral variation of DOP of light. 

\section{ Result \& discussion\label{resutl and discussion}}

\begin{figure}[htb!]
    \centering
    \includegraphics[scale=0.5]{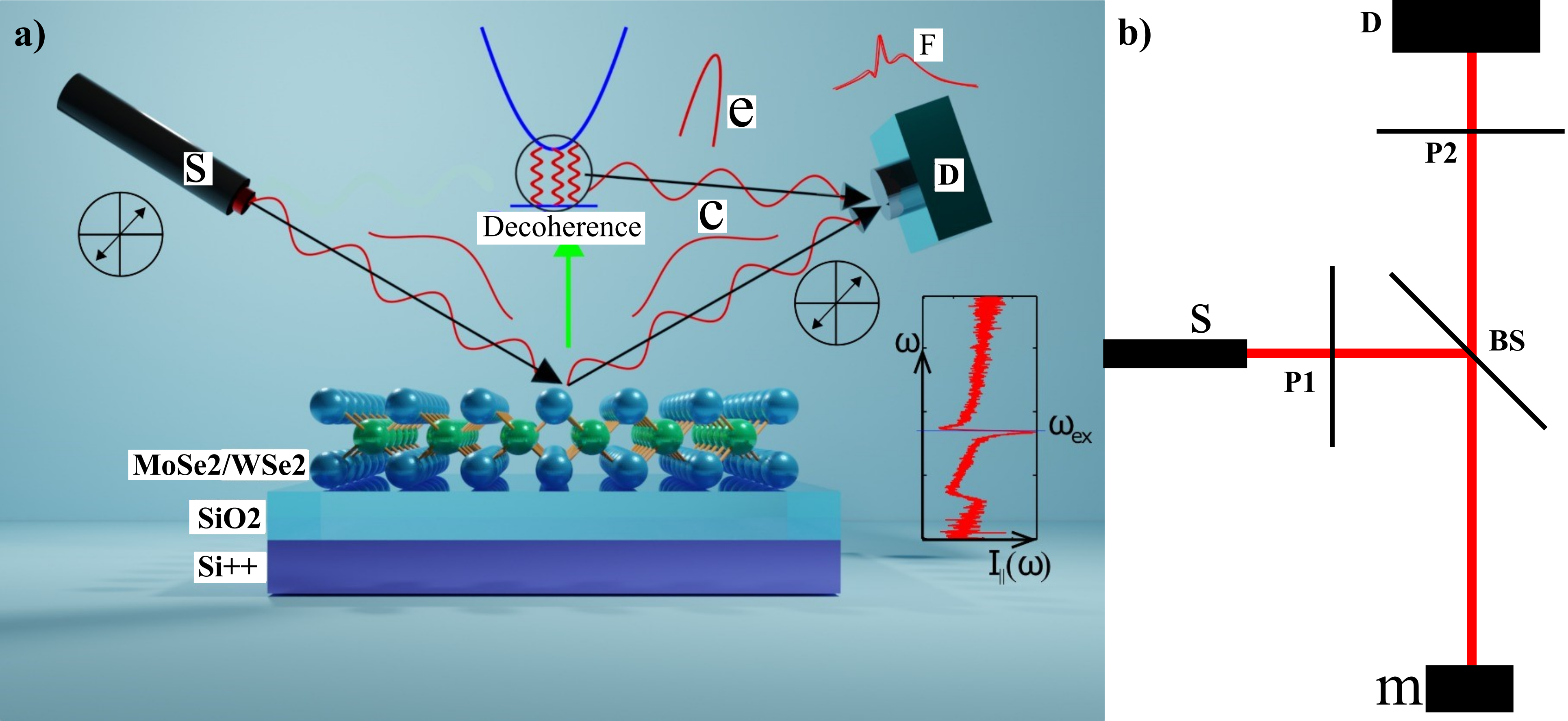}
    \caption{\textbf{Illustration of the emergence of Fano resonance in the reflectance spectra of $MoSe_2$(or $WSe_2$)} (a) The $MoSe_2$(or $WSe_2$) monolayer is exfoliated on Silicon dioxide (SiO2)/ Silicon (Si++) substrate. The spectral domain interference between the narrow excitonic mode (shown as $e$) of $MoSe_2$ and the broadband reflection background (shown as $c$) yields a Fano-type asymmetric spectral line shape $(F)$ in the reflectance spectra. A typical experimental spectrum $(I(\omega)$) showing the sharp excitonic mode and asymmetric line shape is shown in the inset. The direction of the input diagonal polarization $(P: +45^\circ polarized)$ and the detected polarizations are marked. $S:$ light source, $D:$ detector. (b) A schematic diagram of the experimental setup. $P1$ is the input polarizer to generate the polarized input state of the light. $P2$ is the output polarizer to project the reflected light from the sample to a specific polarize state, BS is the beam splitter, and $m$ is the monolayer TMD sample.}
    \label{fig:figure 1}
\end{figure}

For the experimental demonstration of the concept, we choose a system of monolayers of Transition Metal Dichalcogenides (TMD) : $MoSe_2$ and $WSe_2$, which exhibited Fano resonance in the reflection spectra. We observed the asymmetric spectral line shape of Fano resonance separately in the reflected spectra from both samples exfoliated on a dielectric substrate. It is well-known that in such a system, optical fields get coupled with two-dimensional excitons. (Figure.\ref{fig:figure 1}).
Fano resonance in this system emerges due to the spectral domain interference of the corresponding spectrally narrow excitonic mode with a broad continuum of background reflection. Note that in the  Fano interference in this kind of excitonic system, there is an additional mechanism of coherence which is the so-called valley coherence\cite{valley.qiu2019room}. While this valley coherence of the excitonic system may also influence the overall degree of coherence of the spectral domain Fano interference, our primary target was to understand the influence of the overall coherence on the resulting spectral variation of DOP. Thus, even though it may be possible to separate out the contribution of the valley coherence alone from the observed degree of coherence of spectral domain Fano interference, this was not attempted in this study. We recorded both the polarization blind and polarization-resolved reflectance spectra from the systems. For the polarization-resolved measurements, we used P (diagonal: $+45^\circ$) polarized light from a broadband light source (Thorlabs SLS201L) incident perpendicularly on the TMD monolayer. The reflected light from the sample was passed through another linear polarizer (P2) before spectrally resolved signal detection in a spectrometer (Princeton Instruments SP-2750). The reflectance spectra were measured for four different angles of orientation of the analyzes (P2) (Figure. \ref{fig:figure 1}) - along the  direction of the incident diagonal polarization (P), orthogonal to the incident polarization (M - anti-diagonal: $-45^\circ$), along the horizontal direction (H: $0^\circ$), along the vertical direction (V: - $90^\circ$), respectively. The spectral variation of DOP of the reflectance spectra was determined using these measurements. The experimental results are presented subsequently.

\subsection{Emergence of Fano resonance in reflectance spectra of TMD and the role of coherence on the asymmetric spectral line shape}
\begin{figure}[t!]
    \centering
    \includegraphics[scale=0.7]{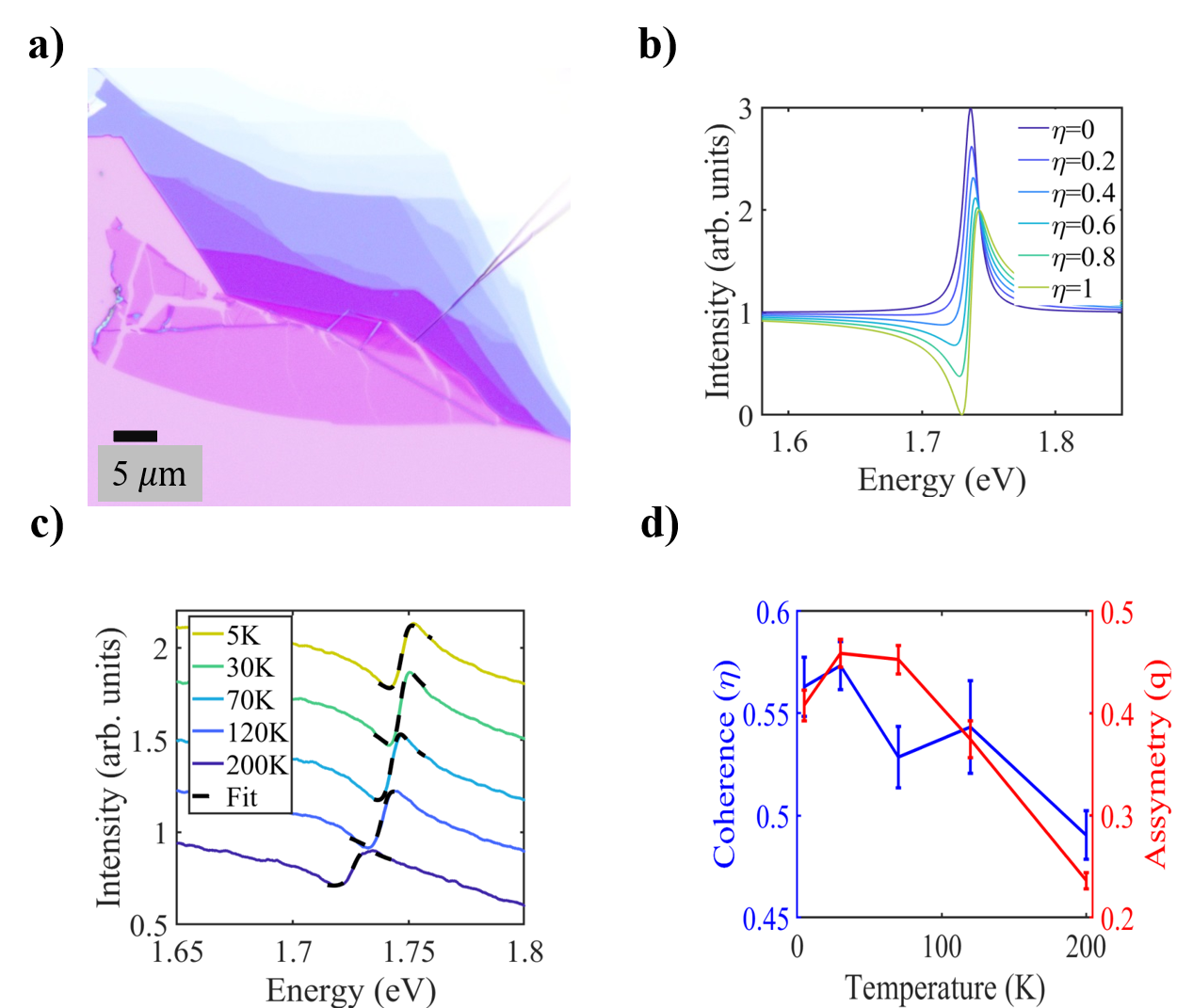}
    \caption{\textbf{Manifestation of the degree of coherence of the spectral domain interference of the modes in the asymmetric spectral line shape of Fano resonance.} ) (a) Microscopic image of the monolayer of $WSe_2$ placed on Silicon dioxide $(SiO_2)$/ Silicon $(Si)$ substrate. (b) Theoretically generated  (using Eq. \ref{eqs: Fano inensity eta}) Fano spectral line shape for varying degrees of coherence $(\eta)$ ($\eta$ between $0$ to $1$ in steps of $0.2$) between the interfering discrete and the continuum modes. A decrease in the magnitude of $\eta$ reduces the spectral dip or the contrast of the spectral-domain interference. (c) Reflectance spectra of $WSe_2$ at five different temperatures ($5K$ to $200K$) and the corresponding theoretical fit (black dashed line) with Eq. \eqref{eqs: Fano inensity eta}  (d) The variations of the corresponding fitted values of $\eta$ (left axis blue line) and the Fano asymmetry parameter $(q)$ (right axis red line) of $WSe_2$ as a function of temperature. }
    \label{fig:figure 2}
\end{figure}
The variation of Fano spectral line shape with changing the degree of coherence $(\eta)$ is illustrated in Figure \ref{fig:figure 2}. The actual image of the $WSe_2$ sample used in our study is shown in Figure \ref{fig:figure 2}a. The theoretical results (using Eq.\eqref{eqs: Fano inensity eta}) are shown in figure \ref{fig:figure 2}b for the spectral range ($E = \hbar\omega = $ 1.58 to 1.85 eV). For complete coherence, i.e., $\eta=1$, perfect Fano interference appears and the line shape represents ideal asymmetric spectral Fano resonance (Figure \ref{fig:figure 2}b). On the other hand, for complete decoherence $(\eta=0)$, the Fano interference line-shape becomes a Lorentzian on a broad continuum background (Figure \ref{fig:figure 2}b), as predicted by Eq.\eqref{eqs: Fano inensity eta}. Note that, with decreasing degree of coherence $(\eta)$ the Fano spectral dip or the contrast of the spectral-domain interference and accordingly the asymmetry parameter $q$ also decreases because it deviates from ideal destructive interference. We experimentally observed  Fano resonance in the reflectance spectra of both $MoSe_2$, and $WSe_2$ systems. The dependence of the Fano spectral line shape on the degree of coherence $(\eta)$ for the $WSe_2$ system are summarized in Figure \ref{fig:figure 2}c and Figure \ref{fig:figure 2}d. The variation of the degree of coherence $(\eta)$ in this scenario is achieved by changing the temperature of the system as increasing temperature is known to reduce the degree of coherence $(\eta)$\cite{valley.yan2015valley}. The experimentally observed variations in the Fano spectral line shape (Figure \ref{fig:figure 2}c) are qualitatively similar to the theoretical trends of figure \ref{fig:figure 2}b. Both the values of $\eta$  and the asymmetry parameter $q$, which are determined by fitting the experimental spectra with Eq. \eqref{eqs: Fano inensity eta}, are observed to decrease with increasing temperature suggesting prominent decoherence effects. As expected, a decrease in the coherence with increasing temperature manifested as a decrease in the spectral sharpness around the Fano dip. At the minimum temperature (4K), the $\eta$, $q$ value become near maximum $(\eta\sim0.56, q\sim0.40)$, and at the maximum temperature (200K) $\eta$, $q$  assume minimum values $\eta\sim0.49, q\sim0.23$ (Figure \ref{fig:figure 2}d). Although we observed  a similar trend in both $WSe_2$ and $MoSe_2$ (not shown) samples, one important  difference was that the $WSe_2$ system generally yielded a much higher degree of coherence\cite{valley.qiu2019room}. As previously discussed, in the spectral Fano interference of these TMD systems, the valley coherence of the excitonic mode is also a contributor to the overall degree of coherence. It is known from previous studies that the valley coherence is much stronger in the $WSe_2$ system as compared to the $MoSe_2$ system\cite{valley.qiu2019room,valley.hao2017neutral}. Traditionally, such valley coherence is determined by measuring the degree of linear polarization (DOLP) of photoluminescence from the excitonic modes\cite{valley.qiu2019room}. The results of such measurements are shown in S1 (supporting information), where the DOLP of photoluminescence spectra with excitation at $\lambda_{ex} = 660 nm$ are shown for both the $WSe_2$ and $MoSe_2$ samples. Clearly, the DOLP of photoluminescence for the $MoSe_2$ is much weaker than that for $WSe_2$. The much stronger degree of valley coherence also accordingly led to a much higher degree of polarization (close to unity) of the Fano resonant reflectance spectra of $WSe_2$. Since our aim was to study the spectral variation of the degree of polarization of Fano resonance and its dependence on the net coherence of the spectral domain interference of Fano resonance, we choose the $MoSe_2$ system which exhibited slightly lower DOP as compared to $WSe_2$. This enabled us to clearly distinguish the variation of the DOP across the Fano spectral dip and the peak.      

Having experimentally demonstrated the role of the coherence of spectral domain interference on the Fano resonance spectral line shape, we thus now turn to the main problem on investigating the influence of the coherence of spectral domain interference on the spectral variation of DOP.  

\begin{figure}[t!]
    \centering
    \includegraphics[scale=0.7]{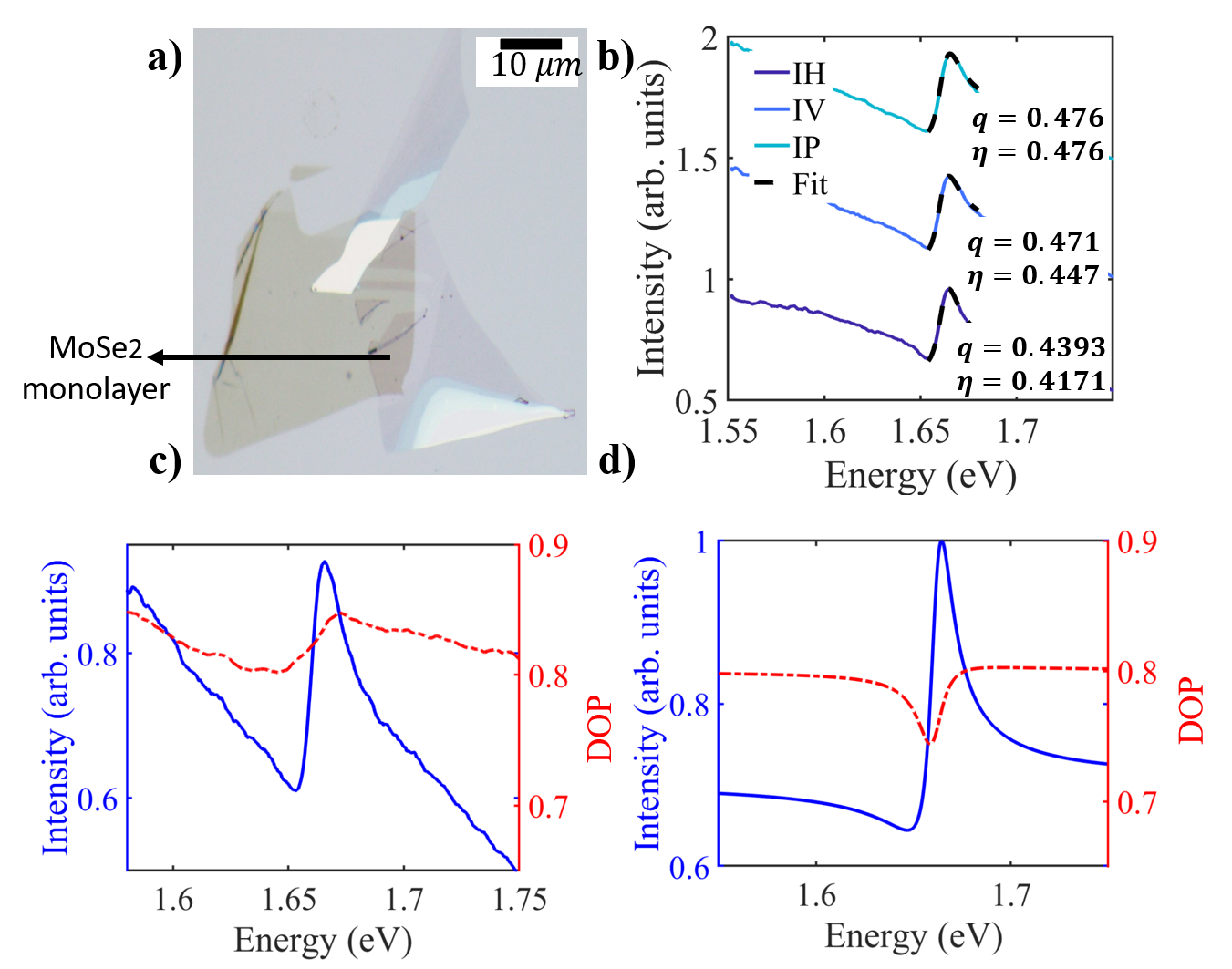}
    \caption{\textbf{Direct evidence of the influence of the degree of coherence of the spectral domain Fano interference on the degree of polarization.} (a) An image of the $MoSe_2$ as exfoliated on a $SiO_2$ substrate used in our study is shown. (b) The experimental Fano reflectance spectra  from the $MoSe_2$ sample with "diagonal" (P) linear polarization of the incident beam. The detected intensities are shown for different post selections of polarization - P (IP), H (horizontal - 0 deg., IH), V (vertical - 90 deg., IV). The spectra are fitted to Eq. \ref{eqs: Fano inensity eta} to obtain the different $\eta$ and $q$  parameters. Note that the $\eta$-parameter determined from IP (representing the 3rd Stokes vector element S3) quantifies the coherence parameter $\eta_{xy}$ in Eq. \ref{eqs:Fano Wxx/yy}. The corresponding $\eta$ determined from  IH and IV represent $\eta_{xx}$ and $\eta_{yy}$. (c) The spectral variation of the degree of polarization (using Eq. \eqref{eqs:dop Stokes}) (right axis red line) and the corresponding Fano line shape (left axis blue line). The variation of DOP follows the spectral interference with its minimum at the Fano spectral dip (corresponding to destructive interference) and maxima at the Fano spectral maxima (constructive interference) as anticipated from Eq. 
    \eqref{eqs:dop W}. (d) The corresponding theoretical variations using Eq. \ref{eqs:dop W}. The input parameters are taken from the experimental fitting and the values are $q=0.4393$, $\eta_{xx}=0.4171$, $\eta_{xy}=0.476$, $\gamma \sim .003$.}
    \label{fig:figure 3}
\end{figure}
The influence of the degree of coherence of spectral domain Fano interference of the two modes on the spectral variation of the degree of polarization of Fano resonance is experimentally demonstrated in Figure \ref{fig:figure 3}. The actual image of the $MoSe_2$ sample used in our study is shown in Figure \ref{fig:figure 3}a. The results are shown for experiments performed at 4K temperature, which was chosen to obtain the maximum possible degree of coherence. The variations of the asymmetric spectral response of Fano resonances in the reflected spectra($E = \hbar\omega=$1.55 eV to 1.75 eV) are shown for P-polarized incident light with H, V, and P polarization post-selections at the detection end (Figure \ref{fig:figure 3}b). Each of the spectral profile of Fano resonance were fitted to Eq.\eqref{eqs: Fano inensity eta} to estimate all the spectral domain Fano interference parameters ($q=0.476$, $\eta=0.476$ for P post-selection; $q=0.471$, $\eta=0.447$ for V post-selection; and $q=0.439$, $\eta=0.417$ for H post-selection). The $\eta$-parameter obtained from the spectra with P-polarization post-selection (which also represents  the 3rd Stokes vector element U) is used to quantify the coherence parameter $\eta_{xy}$ as given in Eq.\eqref{eqs:Fano Wxx/yy}. The corresponding $\eta$ for H-polarization post-selection and V-polarization post-selection represent $\eta_{xx}$ and $\eta_{yy}$, respectively. As discussed previously in the context of the equation connecting coherence and polarization (Eq.\eqref{eqs:dop W}), for physical realizability of the state of polarization of light, $\eta_{xy}>\eta_{xx/yy}$. In conformity with this, we have obtained the values, $\eta_{xy}=0.476$, $\eta_{xx}=0.447$, and $\eta_{yy}=0.417$ (Figure \ref{fig:figure 3}b).  In figure \ref{fig:figure 3}c, we have shown the Fano spectral line shape ($E=\hbar\omega=$1.55 eV to 1.75 eV) and the spectral variation of the DOP, simultaneously. The most noticeable feature of this figure is that the DOP varies as a function of the energy $(E =  \hbar\omega)$. Remarkably, the DOP attains its minimum value $(\sim 0.80)$ near the frequency of the Fano dip corresponding to the destructive interference of the two modes in the Fano resonance and the maximum value of DOP is observed $(\sim 0.85)$ near the frequency of the Fano peak corresponding to constructive interference. These results are in good qualitative agreement with the corresponding theoretical predictions using  Eq. \ref{eqs:dop W} (Figure. \ref{fig:figure 3}d). For the purpose of the theoretical predictions, the Fano interference parameters were obtained by fitting the experimental Fano resonance spectra (figure \ref{fig:figure 3}c) with Eq. \eqref{eqs: Fano inensity eta}, which were subsequently used in Eq. \eqref{eqs:dop W}. Although exact quantitative values are slightly different, the observed trends are similar in the sense that the DOP attains its minimum value near the frequency of the Fano dip, and the maximum value of DOP is obtained near the Fano peak.  The maximum and the minimum values of $DOP=0.80$ and $DOP= 0.75$ are also in good agreement with the corresponding experimental values. These results provide conclusive evidence on the dependence of the degree of polarization of light on the coherence of spectral domain interference of two modes in Fano resonance.  

\section{Conclusions \label{conclusions}}

In summary, we have demonstrated the spectral degree of coherence of the two interfering modes of the Fano resonance system influence the DOP of the emitted light. This is theoretically shown by including a cross-spectral density matrix, including the spectral domain Fano interference. The prediction of this model is experimentally verified on Fano resonance spectral from $MoSe_2$ and $WSe_2$ TMD system. By controlling the temperature of the system, the degree of coherence of the spectral domain interference of Fano resonance was controlled. It was observed that the degree of polarization of light spectrally varies across the dip and the peak of Fano resonance, providing an exclusive signature of the connection between the degree of polarization and the coherence of the spectral domain Fano-type interference of a spectrally narrow mode and a broad continuum. The experimental results were found to be in good agreement with the corresponding theoretical treatment which is  developed by combining a general electromagnetic model of partially coherent interference of two modes in Fano resonance and the cross-spectral density matrix of the interfering polarized fields of light. We emphasize that there are a number of intriguing optical phenomena that arise from the fine interference effects of electromagnetic waves and electromagnetic modes. We anticipate that such an intriguing manifestation of the connection between the spectral degree of coherence and the DOP will also be manifested in broad variety of such non-trivial optical phenomena involving the interference effects. In the context of the TMD systems, our findings also open up an interesting avenue of studying the valley coherence through the DOP of the scattered light. In general, the established connection between the spectral degree of coherence and the DOP may also have useful applications.
\section{Data Availability Statement}
\label{dataavail}
The data that support the findings of this study are available upon reasonable request from the authors.

\section{Acknowledgement}
\label{acknow}
The authors thank the support of the Indian Institute of Science Education and Research Kolkata(IISER-K), Ministry of Education, Government of India. The authors would like to acknowledge the Science and Engineering Research Board (SERB), Government of India, for the funding (grant No. CRG/2019/005558). SG \& DC additionally acknowledges CSIR, Government of India, for research fellowships.

\section{Conflict of Interest}
\label{conflict}
The authors declare no conflict of interest.
\newpage
\printbibliography
\end{document}